\documentclass[aps,prd,twocolumn,showpacs,floatfix,preprintnumbers,amsfont,amsmath,amssymb,nofootinbib,superscriptaddress]{revtex4-1}%linenumbers

\usepackage{aas_macros}
\usepackage{graphicx}
\usepackage{dcolumn}
\usepackage{bm}
\usepackage[normalem]{ulem}
\usepackage[usenames,dvipsnames,svgnames,table]{xcolor}

\usepackage{xcolor}

\definecolor{colorLink}{rgb}{0.9,0,0} % red
\definecolor{colorCite}{rgb}{0,0.7,0} % green
\definecolor{colorURL} {rgb}{0,0,0.8} % navy
\usepackage[colorlinks=true,linktocpage=true,linkcolor=colorLink,citecolor=colorCite,urlcolor=colorURL]{hyperref}
\usepackage{bbold}
\usepackage{multirow}
\usepackage{fontawesome}
\usepackage{algorithm2e}
\usepackage{algpseudocode}

\newcommand{\be}{\begin{equation}}
\newcommand{\ee}{\end{equation}}

\newcommand{\sk}[1]{}

\def\chieff{\chi_{\rm eff}}

\def\pastro{p_{\rm astro}}

\DeclareMathSymbol{\mhyphen}{\mathord}{AMSa}{"39}

\graphicspath{{./}{figs/}}

\begin{document}

% \title{AGENT: Associating Glitches with Extended Noisy Times}
% \title{Improving gravitational wave search sensitivity with AGENT: Association of Glitches with Extended Noisy Times}
\title{Improving gravitational wave search sensitivity with \texttt{TIER}:\\Trigger Inference using Extended strain Representation}
% \title{Enhancing a matched-filtering search pipeline with machine learning}

 \author{Digvijay Wadekar}
  \email{jayw@jhu.edu}
\affiliation{\mbox{Department of Physics and Astronomy, Johns Hopkins University,
3400 N. Charles Street, Baltimore, Maryland, 21218, USA}}
\author{Arush Pimpalkar}
\affiliation{\mbox{Department of Physics and Astronomy, Johns Hopkins University,
3400 N. Charles Street, Baltimore, Maryland, 21218, USA}}
\affiliation{\mbox{Department of Electronics and Communication Engineering, National Institute of Technology, Tiruchirappalli 620015, India}}

\author{Mark Ho-Yeuk Cheung}
\affiliation{\mbox{Department of Physics and Astronomy, Johns Hopkins University,
3400 N. Charles Street, Baltimore, Maryland, 21218, USA}}
\author{Benjamin Wandelt}
\affiliation{\mbox{Department of Physics and Astronomy, Johns Hopkins University,
3400 N. Charles Street, Baltimore, Maryland, 21218, USA}}
\author{Emanuele Berti}
\affiliation{\mbox{Department of Physics and Astronomy, Johns Hopkins University,
3400 N. Charles Street, Baltimore, Maryland, 21218, USA}}
\author{Ajit Kumar Mehta}
\affiliation{\mbox{Department of Physics, University of California at Santa Barbara, Santa Barbara, CA 93106, USA}}
\author{Tejaswi Venumadhav}
\affiliation{\mbox{Department of Physics, University of California at Santa Barbara, Santa Barbara, CA 93106, USA}}
\affiliation{\mbox{International Centre for Theoretical Sciences, Tata Institute of Fundamental Research, Bangalore 560089, India}}
 \author{Javier Roulet}
\affiliation{TAPIR, Walter Burke Institute for Theoretical Physics, California Institute of Technology, Pasadena, CA 91125, USA}
\author{Tousif Islam}
\affiliation{\mbox{Department of Physics, University of California at Santa Barbara, Santa Barbara, CA 93106, USA}}
\author{Barak Zackay}
\affiliation{\mbox{Department of Particle Physics \& Astrophysics, Weizmann Institute of Science, Rehovot 76100, Israel}}
\author{Jonathan Mushkin}
\affiliation{\mbox{Department of Particle Physics \& Astrophysics, Weizmann Institute of Science, Rehovot 76100, Israel}}
\author{Matias Zaldarriaga}
\affiliation{\mbox{School of Natural Sciences, Institute for Advanced Study, 1 Einstein Drive, Princeton, NJ 08540, USA}}

 \date{May 15, 2024}

\begin{abstract}

We introduce a machine learning (ML) framework called \texttt{TIER} for improving the sensitivity of gravitational wave search pipelines. Typically, search pipelines only use a small region of strain data in the vicinity of a candidate signal to construct the detection statistic. However, extended strain data ($\sim 10$ s) in the candidate's vicinity can also carry valuable complementary information.
We show that this information can be efficiently captured by ML classifier models trained on sparse summary representation/features of the extended data.
Our framework is easy to train and can be used with already existing candidates from any search pipeline, and without requiring expensive injection campaigns. 
Furthermore, the output of our model can be easily integrated into the detection statistic of a search pipeline. Using \texttt{TIER} on triggers from the \texttt{IAS-HM} pipeline, we find up to $\sim 20\%$ improvement in sensitive volume time in LIGO-Virgo-Kagra O3 data, with improvements concentrated in regions of high masses and unequal mass ratios. Applying our framework increases the significance of several near-threshold gravitational-wave candidates, especially in the pair-instability mass gap and intermediate-mass black hole (IMBH) ranges.
\href{https://github.com/JayWadekar/TIER_GW}{\faGithub}

%% Older version
% \textbf{We introduce a machine learning (ML) framework called \texttt{TIER} for improving sensitivity of gravitational wave search pipelines. Typically, search pipelines only use gravitational-wave strain data close to the candidate time ($\sim 0.1$ s) in the detection statistic used to determine significance of a candidate \teja{for low mass events/BNS, we can use a lot more data than 0.1 s... Maybe this can be phrased to convey something like searches typically use only data that contains h(t) while this paper uses more.}. However, extended strain data ($\gtrsim 10$ s) can also carry valuable complementary information. 
% We show that this information can be efficiently captured by ML classifier models trained on sparse summary representation/features of the extended data.
% Our framework is easy to train and can be used with already-existing candidates from any search pipeline, and without requiring expensive injection campaigns. 
% Furthermore, the output of our model can be easily integrated into the detection statistic of a search pipeline. Using \texttt{TIER} on triggers from the \texttt{IAS-HM} pipeline, we find up to $\sim$ 20\% improvement in sensitive volume time in LIGO-Virgo-Kagra O3 data, with improvements concentrated in regions of high masses and unequal mass ratios. Our framework also helps increase the significance of some of the near-threshold gravitational-wave candidates, especially in the pair-instability mass gap and intermediate-mass black hole (IMBH) ranges.
% \href{https://github.com/JayWadekar/TIER_GW}{\faGithub}}

\end{abstract}
\maketitle

\section{Introduction}

The first three observing runs of the LIGO--Virgo--Kagra collaboration (LVK) have yielded nearly 100 binary black hole (BBH) merger detections. These detections come from LVK search pipelines~\cite{O1catalog_LVC2016, gwtc1_o2catalog_LVC2018, lvc_o3a_gwtc2_catalog_2021, lvc_o3a_deep_gwtc2_1_update_2021, lvc_gwtc3_o3_ab_catalog_2021, LVK_O3_IMBH_search, Kum24} and also independent search pipelines~\cite{Wad23_HM_Events,ias_pipeline_o1_catalog_new_search_prd2019, Will24_Outside_LVK_catalog_review, ias_o2_pipeline_new_events_prd2020, Ols22_ias_o3a, NitzCatalog_1-OGC_o1_2018, NitzCatalog_2-OGC_o2_2020, nitz_o3a_3ogc_catalog_2021, nitz_4ogc_o3_ab_catalog_2021,Chi23,Meh23_ias_o3b, Kol24_Ares_ML_Search} analyzing publicly available data from the Gravitational Wave Open Science Center (GWOSC)~\cite{GWOSC}. To claim a significant detection, the search pipelines have to quantify its false-alarm rate (FAR) and the probability of the signal being of astrophysical origin ($\pastro$). 

\begin{figure}
  \centering
  \includegraphics[scale=0.195,keepaspectratio=true]{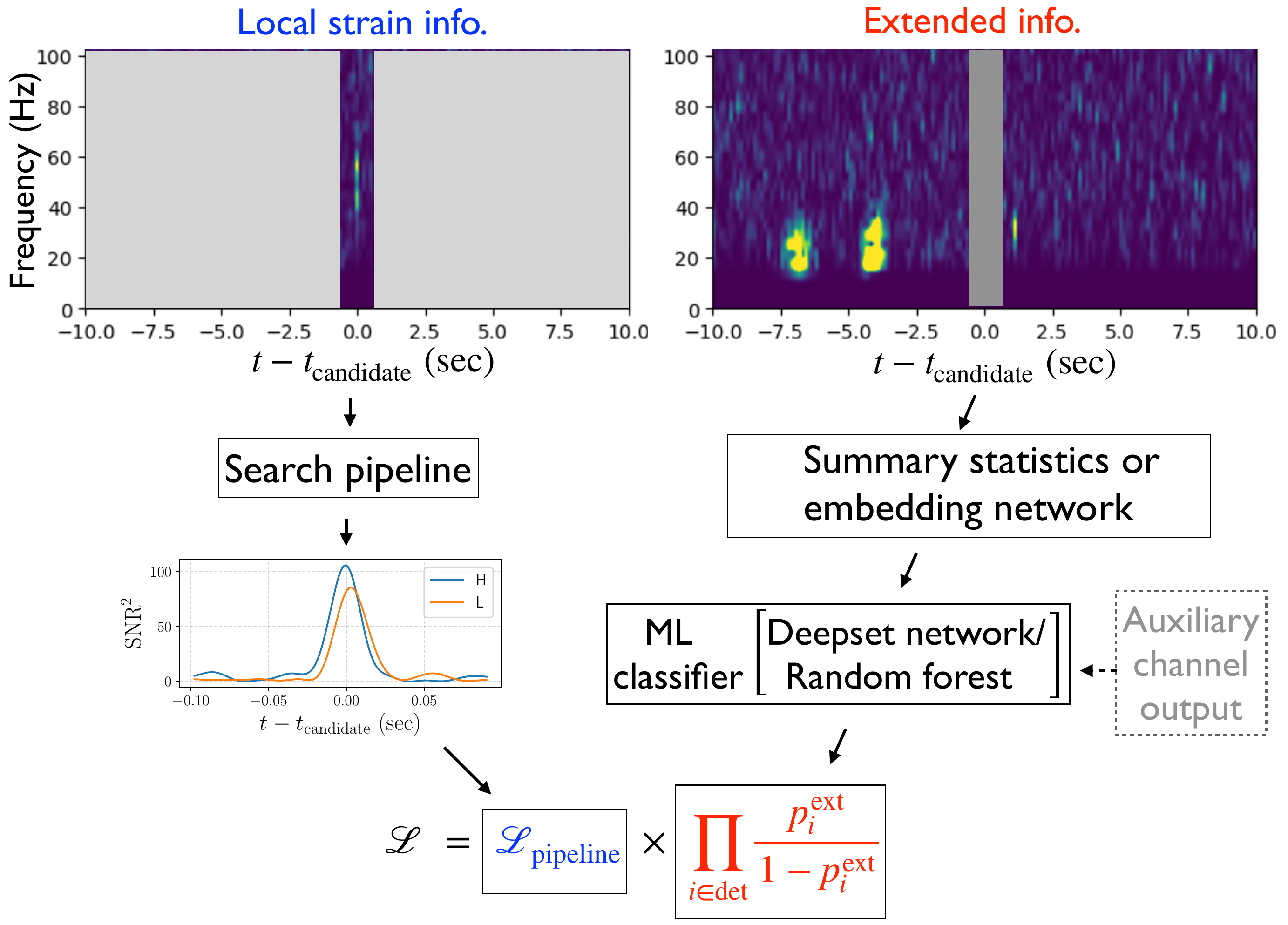}
  \caption{The ranking statistic of a search pipeline ($\mathcal{L}_\mathrm{pipeline}$) uses local SNR to sort or rank a candidate event by its likelihood of being an astrophysical signal rather than noise. The \texttt{TIER} framework can be used to augment the ranking statistic of a pipeline by including complementary information about the extended strain data next to the candidate time (see Eq.~\eqref{eq:NeymanPearson}). To reduce the dimensionality of the extended data, we use a set of sparse summary representation/features as an input to a ML model. We thus obtain the probability that the candidate is an astrophysical signal ($p^\mathrm{ext}$) based on the extended strain data. Auxiliary channel data (e.g., seismometer outputs) can also be included in our ML framework, but this is left to a future study.}
  \label{fig:TIER}
  \end{figure}

For low-mass compact object mergers, the signal has a long duration and a characteristic chirp feature, which cannot be easily mimicked by non-Gaussian transients. Thus, the search background is dominated by Gaussian noise in this regime. On the other hand, high-mass BBHs ($M_\mathrm{total}\gtrsim 100\, M_\odot$) merge at frequencies at the edge of the current detector band, and thus we can only see a few gravitational wave (GW) cycles in band~\cite{lvc_properties_GW190521}. 
Non-Gaussian transients (also called detector ``glitches") dominate the background in this regime as they can easily mimic the short signals~\cite{Son24_O4a_DetChar, BlipGlitches, Son21_Scattered_Light}, making the search for short signals much more challenging~\cite{Che25_IMRI_search}. However, finding high-mass events is crucial, as their astrophysical distribution is currently very poorly constrained. 
Finding such high-mass events can give us insights into pair-instability supernova physics~\cite{Marchant:2020haw, Farmer:2019jed, VanSon20_UMG_Pollution, Mehta:2021fgz, Hen23_UMG, Gol23_UMG}, hierarchical BH formation in star clusters~\cite{umg_hierarchical_must_have_BHspin_Gerosa2021, Fra18_IMBH_LISA, Kri23_NSC, Kri24_GasAccretion, Kri24_Rapster} and active galactic nuclei~\cite{hierarchical_mergers_agn_kocsis2019,agn_bbh_population_chieff_q_simulation_mckernan_ford2019, bbh_spin_evolution_agn_Tagawa_2020a, bbh_evolution_agn_merger_timescale_ishibashi2020a, migration_traps_spins_rates_mckernan_ford2020a, 
mass_gap_agn_bbh_mergers2021, agn_accretion_disk_merger_population2020a, Tag20_AGN, Sam22_AGN}, and phenomenology of population III stars~\cite{Fra24_UMG}.

We currently do not have a good theoretical framework for non-Gaussian noise, but we can use machine learning (ML) tools to learn the distribution of non-Gaussian noise from the detector data and use them to down-weight non-Gaussian transients~\cite{Bis13_ML_noise,GravitySpy,Cuo24_ML_GW_review, Fer25_CNN, Sha22, Sal24_Deepclean, Rei25_Deepclean_coherence, Alvarez-Lopez_GSpyNet, Lop25_ML_noise, Mal25_PE_glitch, Leg24_PE_glitch, 2023arXiv230613787S, Mar25_Aframe_ML, Sch22, McL24_ML_search_BNS, Kap17_Classifier, Kol24_Ares_ML_Search, Ess20_iDQ}. In the limit of infinite data and computational resources, ML methods are optimal, however this idealistic situation is not often realized in practice. On the other hand, combining ML methods with principled matched-filtering
searches can help building tools which are more robust (with respect to
generalizability, interpretability) and can be trained with less data. In this paper, we therefore use ML tools as a complementary background reduction layer on the top of existing methods in our pipeline.

We propose a framework called \texttt{TIER} (Trigger Inference using Extended strain Representation) which can be used to augment the ranking statistic of any given search pipeline (see Fig.~\ref{fig:TIER}). 
Search pipelines typically only look at strain regions close to a candidate ($|t-t_\mathrm{candidate}|\lesssim 0.1$\,s for high-mass BH mergers) when determining a candidate's significance. However, we will show that there is valuable complementary information in the extended strain data ($0.1\,\text{s} \lesssim|t-t_\mathrm{candidate}|\lesssim 10$\,s). We will show that a significant fraction of glitches are more likely to live in noisy environments. Furthermore, glitches also typically cluster together (colloquially referred to as: glitches come with friends). We will show how ML algorithms can be used to efficiently capture this extended strain information and improve our search sensitivity. The ML classifier model learns the correlation between the nature of a candidate and the extended strain data surrounding it. Thus, it can be used to probabilistically downweight triggers which live in noisy environments.
% One empirical observation which the \texttt{TIER} framework exploits is that, unlike real gravitational wave events, glitches or non-Gaussian transients cluster in time. Also, glitches have a higher chance of living in relatively noisier environments.

We first discuss how to include the extended strain information in the search ranking statistic in Section~\ref{sec:ranking_statistic}. We then discuss the data used to train the ML models and the summary statistics used to compress the extended strain information in Section~\ref{sec:data}. We then present the ML models we used in Section~\ref{sec:ML_models}, and the results obtained in Section~\ref{sec:results}. We finally discuss the implications of our results in Section~\ref{sec:discussion}, and conclude in Section~\ref{sec:conclusion}.

% We empirically see that glitches are more likely to live in noisy environments and vice versa. We will show how ML algorithms can be used to accurately capture this environmental information and improve our search sensitivity.

%In the infinite data and compute limit, ML methods are optimal, however this idealistic situation is not often realized in practice. 
%Combining ML methods with principled matched-filtering
%searches can help building tools which are more robust (w.r.t
%generalizability, interpretability) and can be trained with less data.

% This makes a common lore: glitches come with friends nearby. We aim to make this quantify this effect and use it to improve our search sensitivity.
% Searches typically only look at data regions close to a trigger ($t-t_\mathrm{trigger}\lesssim 0.1$ s). However, we will show that there is valuable information in the extended environmental regions ($t-t_\mathrm{trigger}\lesssim 10$ s) and ML algorithms can use this environmental information and help us improve our search sensitivity.

\section{Including environmental information in the search ranking statistic}
\label{sec:ranking_statistic}

To compute the significance of candidates (i.e., their FAR), one typically uses a ranking statistic ($R$). This helps us compare the significance of foreground candidates (typically obtained by coincident analysis of multiple detector datastreams) against the background triggers (typically obtained from the timeslides method, see Fig.~3 in~\cite{Dav20_PyCBC_3det} for a schematic illustration). The FAR of a foreground coincident candidate is determined by the number of background triggers whose ranking statistic is larger than that of the candidate (FAR$_\mathrm{candidate} \propto N (R_\mathrm{backgrounds}>R_\mathrm{candidate})$).
The optimal ranking statistic is given by the Neyman--Pearson lemma~\cite{neymanpearson} as the ratio of Bayesian evidence of the candidate under the signal ($\mathcal{S}$) and the noise ($\mathcal{N}$) hypothesis. This can be written to include the nonlocal/extended strain information as:
%\markc{If we want to be extra precise, perhaps we should say that $\mathcal{S}$ here means that $d_\mathrm{local}$ triggered a template in a given bank with rank $R >$ some threshold, and $\mathcal{N}$ is the same but for noise.}

  \begin{subequations} \label{eq:NeymanPearson}
    \begin{align}
    \mathcal{L} &\equiv \frac{P(d|\mathcal{S})}{P(d|\mathcal{N})}
    \equiv \frac{P(d_\mathrm{local}, d_\mathrm{nonlocal}|\mathcal{S})}
    {P(d_\mathrm{local}, d_\mathrm{nonlocal}|\mathcal{N})} \label{eq:NeymanPearson1} \\
    &=  \frac{P(d_\mathrm{local}|\mathcal{S})}{P(d_\mathrm{local}|\mathcal{N})}
    \frac{P(d_\mathrm{nonlocal}|d_\mathrm{local},\mathcal{S})}
    {P(d_\mathrm{nonlocal}|d_\mathrm{local},\mathcal{N})} \label{eq:NeymanPearson2} \\
    &\simeq \mathcal{L}_\mathrm{pipeline} \prod_{i\in \mathrm{det}} \frac{p^\mathrm{ext}_i}{1-p^\mathrm{ext}_i} \label{eq:NeymanPearson3}
    \end{align}
    \end{subequations}
where $\mathcal{L}_\mathrm{pipeline}$ is the traditional evidence ratio used by search pipelines as their ranking statistic. Search pipelines have traditionally only used strain information close to the candidate time $d_\mathrm{local}$ (e.g., within ${}\pm0.1$\,s in the previous IAS pipeline studies). The second term in Eq.~\eqref{eq:NeymanPearson2} corresponds to the evidence ratio obtained by using nonlocal information. However, it is challenging to model $P(d_\mathrm{nonlocal}|d_\mathrm{local},\mathcal{N})$ analytically, as we do not currently have a good theoretical framework for non-Gaussian noise. ML methods circumvent this by learning a model directly based on training examples, as detailed in the next section. We will also show that ML methods provide the probability that the candidate is an astrophysical signal ($p^\mathrm{ext}_i$) based on $d_\mathrm{nonlocal}$ in the $i^\mathrm{th}$ detector. This can directly be used in the ranking statistic as per Eq.~\eqref{eq:NeymanPearson3}. Note that in Eq.~\eqref{eq:NeymanPearson2}, in the case of stationary Gaussian noise, the density ratio of nonlocal data reduces to 1, as expected.

%Note that we have made the assumption $P(d_\mathrm{local}, d_\mathrm{nonlocal}|\mathcal{N}) = P(d_\mathrm{local}|\mathcal{N})P(d_\mathrm{nonlocal}|\mathcal{N})$ in Eq.(2.b) (i.e., the nonlocal and local information are independent). This assumption helps us make the ranking statistic modular. We leave the exploration of the more general case to a future study.

%%%%%%%%%%%%%%%%%%%%%%%%%%%%%%%%%%%%%%%%%%%%%%%%%%%%
\section{Training data for ML model}
\label{sec:data}

We use the data output from the \texttt{IAS-HM}\footnote{\label{foot}The $\tt{IAS-HM}$ pipeline is publicly available at \url{https://github.com/JayWadekar/gwIAS-HM}} O3 search in \citet{Wad23_HM_Events} to train and test our ML models. Although the \texttt{IAS-HM} search included the higher-order modes (HM), the \texttt{TIER} framework could as easily be used with quadrupole-only search pipeline outputs. We use the whitened data from the search, which has been processed to remove loud noise disturbances (we also use inpainting to precisely remove the contribution of such noisy regions during matched filtering~\cite{psd_drift}). Details of the template banks used are given in~\cite{Wad23_TemplateBanks},
%(it is worth mentioning that we also use ML tools to improve the efficiency of GW template banks~\cite{Wad23_TemplateBanks}).
and the trigger collection algorithm is presented in~\cite{Wad23_Pipeline}.

\subsection{Background and signal candidates}
\label{sec:Bkg_signal_candidates}

To train the ML model, we first need to collect the $d_\mathrm{nonlocal}$ samples corresponding to the signal and noise hypotheses to learn the distributions contributing to the evidence ratio in the RHS of Eq.~\eqref{eq:NeymanPearson2}:

$(i)\, P(d_\mathrm{nonlocal}|d_\mathrm{local},\mathcal{N})$:  To learn this distribution, we use the background triggers already collected in the search pipeline by the method of timeslides. In the timeslide method, we shift the strain timeseries from one of the detectors by times greater than the light-crossing time between the detectors ($t_\mathrm{shift}\geq 0.1$\,s) and redo the coincidence step of the pipeline on the shifted strains. We perform 2000 such runs for each of O3a and O3b. This helps us collect background triggers (which are essentially noise fluctuations in individual detectors which spuriously align with each other upon time-shifting). We only use the background triggers which cleared all the veto tests and threshold cuts set in the pipeline.
%\markc{I think what we were learning is $ P(d_\mathrm{nonlocal}|\mathcal{N})$ instead of $P(d_\mathrm{nonlocal}|d_\mathrm{local},\mathcal{N})$, so it does not care at all what is happening in $d_\mathrm{local}$ right? Basically we are assuming that $d_\mathrm{local}$ and $d_\mathrm{nonlocal}$ are independent right? If that's the case maybe we should discuss it.}
%\markc{The only thing we care about in $d_\mathrm{local}$ is that it triggered a template in the given bank, and the candidate has a rank above some threshold, but this seems to be implied by the $\mathcal{N}$ already (see my comment before Eq.~\eqref{eq:NeymanPearson}).}
%\markc{Same comment for the $(ii)\, P(d_\mathrm{nonlocal}|d_\mathrm{local},\mathcal{S}):$ below.}

Note that the background triggers were already collected for estimating the FAR of coincident candidates.  We reuse the same background triggers to collect $d_\mathrm{nonlocal}$ samples corresponding to their times. We choose to use $d_\mathrm{nonlocal}$ samples corresponding to the top 20,000 background triggers for each bank to train the ML model. These triggers have the highest ranking statistic scores and contribute to the estimation of significance of coincident triggers with $\mathrm{IFAR}\gtrsim 0.1\,\mathrm y$. In the high-mass banks, the top background triggers mostly correspond to glitches (instead of being due to Gaussian noise, as is the case for low-mass banks). Note that we are currently assuming that the distribution $P(d_\mathrm{nonlocal}|d_\mathrm{local},\mathcal{N})$ does not vary with the particular type or SNR of the glitch present in $d_\mathrm{local}$ (for the chosen set of top background triggers). This assumption does not bias our results, but can make our method less effective. A more accurate way could be to instead model the $d_\mathrm{nonlocal}$ distribution being conditional on the properties of the glitch in $d_\mathrm{local}$: $P(d_\mathrm{nonlocal}|\mathrm{glitch\, type}, \mathrm{trigger\, SNR,...})$, but we leave this exploration to a future study. %we discuss this point further in section~\ref{sec:Future_work}

$(ii)\, P(d_\mathrm{nonlocal}|d_\mathrm{local},\mathcal{S}):$  To learn this distribution, we first assume that GW mergers follow a Poisson process with rate fixed throughout the run (unlike the glitches, the GW mergers are not expected to artificially cluster at particular times). Ideally, one could do an expensive injection-recovery campaign with injection times sampled uniformly across the observing run. Similar to the background case, one can collect $d_\mathrm{nonlocal}$ samples corresponding to the times of the top injections which were detected by our pipeline. Note that the extent of $d_\mathrm{local}$ is chosen such that the injected GW waveform is safely contained within $d_\mathrm{local}$ and does not enter $d_\mathrm{nonlocal}$. 

We are only interested in collecting $d_\mathrm{nonlocal}$ samples in this paper, and thus all the information being used from a potential injection campaign is the set of times of the injections. We instead create a set of simulated times in a much cheaper but approximate way without using injections. We first sample from a uniform distribution of times between the start and endpoint of the run. Similar to what is done in actual injection campaigns~\cite{lvc_o3a_population_properties_2021}, we remove the hopeless times (i.e., those when a particular detector is off or the data quality is very bad) in individual detectors from our sample. We apply the hopeless cuts in a similar way as we do for actual triggers in the pipeline (e.g. we remove the times occuring within segments flagged for data quality issues by LVK). In Fig.~\ref{fig:PSD_local_distribution} in the Appendix, we show the distribution of properties of simulated times with the times of actual recovered injections (which we later use in section~\ref{sec:results} for comparing pipeline sensitivity). We find that the cheaply simulated set of times have very similar properties to the times of the recovered injections, thus validating our cheaper method. We also find that the properties of background trigger times are significantly different from the injection times. This difference will be later used by our trained ML model to classify signals from background triggers.

    \begin{figure*}
      \centering
      \includegraphics[scale=0.7,keepaspectratio=true]{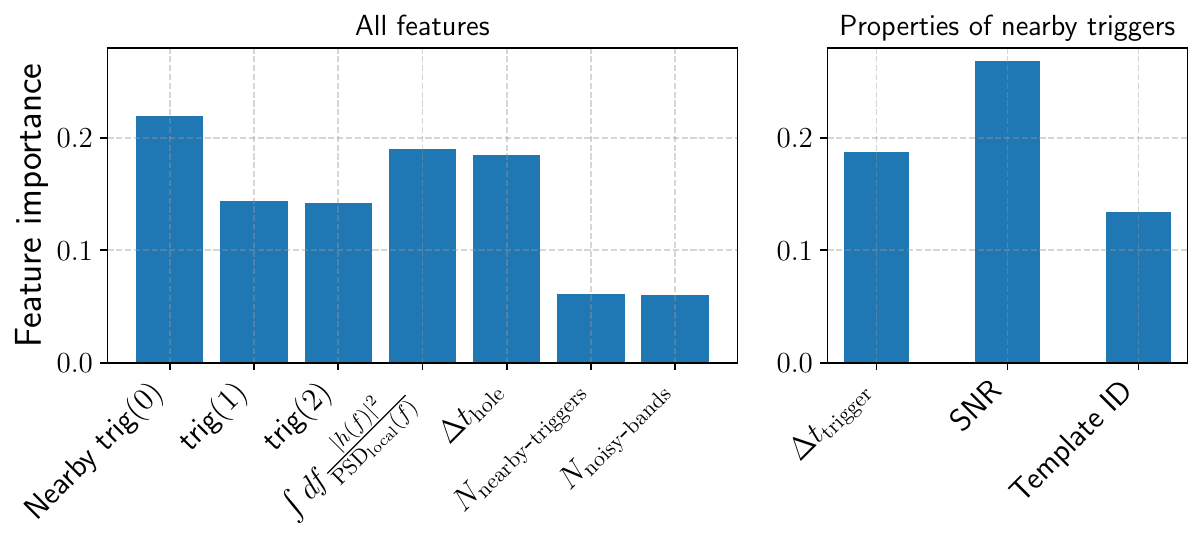}
      \vspace{-0.5cm}
      \caption{We compress the extended strain data into a few summary statistics/features to make the ML models easier to train and more interpretable. Here, we list the summary features (see Section~\ref{sec:SummaryStats} for the feature definitions) and show their importance for the Random forest classifier (RF) prediction for a particular high-mass template bank ($M_\mathrm{tot}\sim 250\, M_\odot$). The three nearby triggers labeled in the plot correspond to the three loudest triggers within $|\Delta t|< 15$\,s, where $\Delta t = t_\mathrm{nearby\mbox{-}trigger}-t_\mathrm{candidate}$. For each of the nearby triggers, we show the importance of the $\Delta t$, the SNR of the trigger, and its template ID in the right panel. We see that the total number of triggers in the extended strain data plays a relatively minor role, whereas the properties of the loudest nearby triggers and the local PSD are the most important.}
      \label{fig:Feature_Imp}
      \end{figure*}

\subsection{Summary statistics for candidate environment}
\label{sec:SummaryStats}

One could think of directly training a ML model on the entire strain data within an extended region as input. This however leads to the input data having a large dimensionality: $\sim 10^4 ( \simeq 10\,\mathrm{s}\times 1024$\,Hz). This makes the ML model expensive to train and more prone to overfitting (if a model with a lot of learnable parameters is used). We will therefore resort to using a few summary statistics/features of the environmental strain in this paper, and leave training the ML algorithm directly on the full-dimensional extended strain data to a separate paper~\cite{Pim25_Env_ML}. 

We describe below the summary statistics used to train the ML models. We also show the importance of these quantities in the ML model predictions in Fig.~\ref{fig:Feature_Imp}. It is important to note that the following summary statistics were developed using intuition and an empirical trial-and-error method. Devising better summary statistics or compressing the environmental data using an ML model could give even better results. This direction will be probed in the upcoming Ref.~\cite{Pim25_Env_ML}. 

\begin{enumerate}
  \item $\Delta t_\mathrm{triggers}$ (separation of nearby triggers from the candidate) -- To analyze a candidate trigger at $t_\mathrm{candidate}$, we make a list of nearby triggers (i.e., those falling within $0.5<|t-t_\mathrm{candidate}| < 15$\,s) from the catalog already collected at each detector in the matched-filtering stage of the search pipeline\footnote{In the IAS pipeline, we first separately collect triggers from individual detectors. The multi-detector coincidence stage of the pipeline is later performed from these collected sets.}. We then compute the relative separation of the triggers from the reference candidate, $\Delta t_i \equiv t_i - t_\mathrm{candidate}$. We use a lower limit of 0.5\,s in the above equation to exclude overlap with candidates originating from long gravitational waveforms (in principle, this value could be precisely tuned according to the length of the waveform). Note that we only consider BBHs with individual masses $>10 M_\odot$, where the signal duration is safely below 0.5\,s.
  %\markc{We are only looking at the stretches of data $d_\mathrm{local}$ that are coincident triggers, but here we are using the single-detector triggers for $\Delta t_\mathrm{triggers}$. Some might wonder why we don't use the coincident triggers also for $\Delta t_\mathrm{triggers}$. It is obvious to us but perhaps we should discuss this a little bit.}
  
  \item SNRs of nearby triggers -- We also have information about the SNR of the nearby triggers in each detector. As our search includes higher harmonics, we use the full harmonic SNR ($|\rho^2_\mathrm{HM}|\equiv|\rho_{22}|^2+|\rho^\perp_{33}|^2+|\rho^\perp_{44}|^2$) of the triggers, where $\rho^\perp_{33}$ and $\rho^\perp_{44}$ correspond to the SNR of the orthogonalized higher-harmonic templates. We use the rescaled quantities version, $\log_{10}(\rho^2_\mathrm{HM})$, to reduce the dynamic range of the inputs to the ML algorithm. We checked that if we alternatively use the quadrupole-only SNR $(\log_{10}|\rho_{22}|^2)$, we obtain similar results. We show a distribution of SNR of the loudest nearby trigger for signals and background candidates in Fig.~\ref{fig:PSD_local_distribution} in the Appendix.
 
 \item Template ID of the nearby triggers -- We use information about the template corresponding to the nearby triggers. In the \texttt{IAS} searches, the templates do not correspond to physical parameters (e.g., $m_1, m_2, s_{1z}, s_{2z}$), but to singular vector components ($c_0, c_1$)~\cite{ias_template_bank_PSD_roulet2019, Wad23_TemplateBanks}. Our template ID therefore corresponds to a unique set of $\{c_0, c_1\}$ values. The logic behind using this term is that glitches are caught more frequently by certain templates (e.g. short-duration waveforms), and providing this information to the ML model could be useful.

  \item $\int\frac{|h_\mathrm{norm}(f)|^2}{\mathrm{PSD}_\mathrm{local}(f)}df$ --  This quantity measures the sensitivity of a given local ($\sim 10$\,s) strain data chunk to gravitational waveforms from a particular template bank. Essentially, this quantity penalizes data chunks with high PSD values (especially for frequencies which are dominant in gravitational waveforms). We show a distribution of the local sensitivity for signals and background candidates in Fig.~\ref{fig:PSD_local_distribution} in the Appendix. The quantity $h_\mathrm{norm}(f)$ corresponds to the normalized amplitude profile of the GW waveforms associated with the bank (the IAS template banks are constructed such that the normalized amplitude profiles are nearly the same for all waveforms within a bank~\cite{Wad23_TemplateBanks}).
  %\MZ{Is this supposed to take into account when astro events are supposed to be more likely? Not sure this is the case if you ``inject" at random times.}

We break the local sensitivity term into two parts:
\begin{equation}
\begin{split}
&\int df \frac{|h_\mathrm{norm}(f)|^2}{\mathrm{PSD}_\mathrm{local}(f)}\\ &\simeq \bigg\langle \frac{\mathrm{PSD}_\mathrm{global}(f)}{\mathrm{PSD}_\mathrm{local}(f)} \bigg\rangle \int df\frac{|h_\mathrm{norm}(f)|^2}{\mathrm{PSD}_\mathrm{global}(f)}\,.
\end{split}\label{eq:local_sensitivity}
\end{equation}
For estimating the second term, we use a ``global'' PSD calculated from strain data in the particular $t_\mathrm{gps}$ file (4096\,s). The first term has been labeled as \emph{PSD drift correction} in Ref.~\cite{psd_drift}. This is a frequency-averaged scalar term accounting for the variation of PSD in relatively short time intervals ($\Delta t \sim 10$\,s), and we compute it using~\cite{psd_drift}:

\begin{equation}
\begin{split}
  \bigg\langle \frac{\mathrm{PSD}_\mathrm{local}(f)}{\mathrm{PSD}_\mathrm{global}(f)} \bigg\rangle^2 &\sim \int_{t - \Delta t / 2}^{t + \Delta t / 2} | h_w(t') \circledast d_w(t') |^2 dt' \\
  &\propto \int \left| \frac{h_\mathrm{norm}^*(f) d_\mathrm{local}(f)}{\mathrm{PSD}_\mathrm{global}(f)} \right|^2 df\,,
\end{split}
\end{equation}
where $h_w(t)$ is the whitened normalized time-domain amplitude profile of the gravitational waveforms associated with the bank, $d_w(t)$ is the whitened strain data in the local ($\Delta t \sim 10$ s) region, and $\circledast$ is the cross-correlation operator. As an input to the ML model, instead of passing the product of the two terms in Eq.~\eqref{eq:local_sensitivity}, we pass the two terms separately. We find that this helps the ML model learn a more optimal combination of the two terms compared to the simple product.

\item $\Delta t_\mathrm{hole}$ -- Distance to nearest ``hole,'' i.e., the regions where data is excised in the \texttt{IAS-HM} search pipeline. These correspond to very loud noise transients and are expected to have SNR $\gtrsim$ 20. We include this parameter separately from the nearby triggers as these data regions do not have an accurate SNR estimate in our pipeline (as we inpaint such loud noisy data segments before estimating the PSD~\cite{psd_drift}).

\item $N_\mathrm{noisy\, bands}$ -- Number of noisy frequency bands/channels overlapping with the trigger. We use results from the \texttt{Band Eraser} tool from~\cite{Wad23_Pipeline} to estimate this quantity. The \texttt{Band Eraser} tool removes individual noisy segments with dimensions $\rm 64\,s \times 2\,Hz$ in the spectrogram of the strain data.

\end{enumerate}

%%%%%%%%%%%%%%%%%%%%%%%%%%%%%%%%%%%%%%%%%%%%%%%%%%%%
\section{Machine learning models used}
\label{sec:ML_models}

\subsection{Random forest classifier (RF)}

In this work, we use the \texttt{RandomForestClassifier} (RF) module from the \texttt{sklearn} package. We use \texttt{gini\_impurity} criterion to make the leaf splits. We also use the \texttt{max\_depth} and \texttt{min\_samples\_split} parameters to control the complexity of the model (in order to reduce overfitting). 

The RF model is quick to train and also provides global importance of the input features based on their relative contribution to reducing the loss (see Fig.~\ref{fig:Feature_Imp}). This enhances the model's interpretability. Let us briefly compare this approach with the interpretability techniques used in neural networks (NN). Gradient-based interpretability techniques in NN like saliency maps can help with instance-level explanation (e.g., why a specific instance was classified). However, saliency maps can be noisy, local to each input and harder to interpret. Feature importance from RF is instead global and more intuitive: it tells us which features matter across the full dataset.

One of the downsides of using RF in our case, however, is that the length of the input features for RF needs to be fixed. There can be an arbitrary number of loud triggers next to a particular candidate, but we had to restrict our analysis to a fixed number of top few triggers to keep our model simple (a more complex model will be more prone to overfitting). We chose to use only the top three nearby triggers ranked by their SNR, but we see from the relative feature importance in Fig.~\ref{fig:Feature_Imp} that including more triggers could possibly help us perform better. We thus devise a different deep NN based algorithm in the next section. Note that there can be cases where there are less than three triggers above our collection threshold (with SNR$^2\gtrsim$ 20) in the extended data; we pad zeros to keep the input feature length fixed in those cases. 

\subsubsection{Calibrating RF output probability}
\label{sec:Calibrating_RF_output_probability}

\begin{figure}
  \centering
  \includegraphics[scale=0.75,keepaspectratio=true]{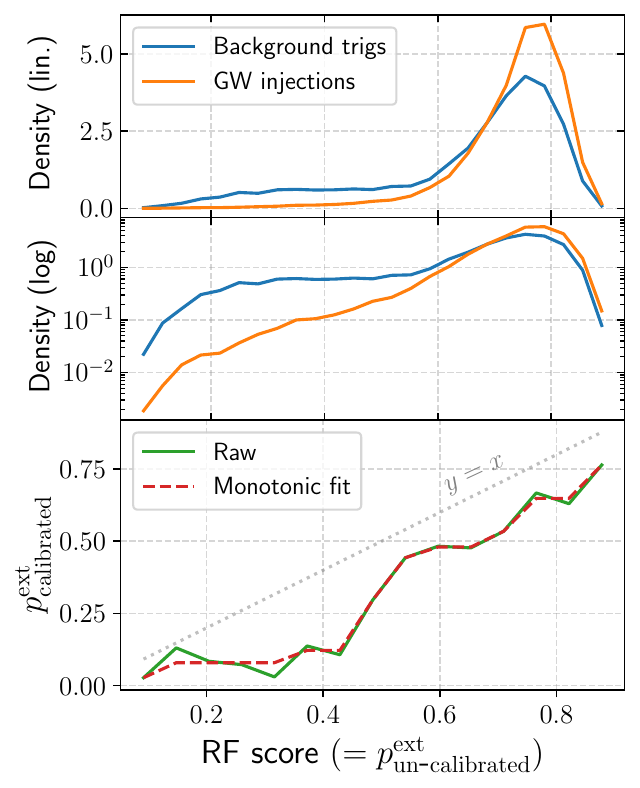}
  \caption{\textbf{Top and center}: After training the ML model (random forest classifier [RF] in this figure), we show the normalized density of score predicted for the two classes in the validation set data. We see that there is a relative overdensity of the GW injections in the high score region, and vice versa for the background triggers. We show the histogram in both log and linear scales for clarity.
  % model can distinguish a subset of glitches from injections based on their environmental information.
   \textbf{Bottom}: It is advisable to calibrate the output probabilities from ML models in order to obtain the true Bayesian probabilities. We show the calibration of the RF output $(p^\mathrm{ext}_\mathrm{uncalibrated})$ using ratio of the histograms in the top panels (see Section~\ref{sec:Calibrating_RF_output_probability} and Eq.~\eqref{eq:Probability_calibration} for details). We perform a monotonic fit to the calibration curve to ameliorate discreteness effects, and then obtain $p^\mathrm{ext}_\mathrm{calibrated}$ values which are used in the rest of the paper.}
  \label{fig:RF_scores}
  \end{figure}

Once the RF is trained, the function \texttt{predict\_proba()}
outputs its classification probability. However, this can be different from the true Bayesian probability, as the RF algorithm can skew the low and high probability tails when there are few training examples. RF consists of decision trees, which are piecewise constant models, not smooth probabilistic models. To correct this, we use a simple calibration algorithm which transforms the classifier output probability  to the Bayesian probability $p(x)$ as:
% \be
% \lim_{\epsilon \rightarrow 0} \int_{r-\epsilon}^{r+\epsilon} dx\, n(x) \propto \epsilon \ \ \ \ \ r \in [0,1]
% \ee
\be
p(x) = \frac{\int_{x-\epsilon}^{x+\epsilon} dx'\, n_\mathrm{signal}(x')}{\int_{x-\epsilon}^{x+\epsilon} dx'\, (n_\mathrm{signal}(x') + n_\mathrm{noise}(x'))}
\label{eq:Probability_calibration}
\ee
where $n(x)$ is the number density of samples in the validation set. The width of the interval $\epsilon$ is chosen to be small enough such that the density is roughly constant within the interval, but also large enough to avoid discreteness effects due to low number of samples. To calculate the density $n(x)$, we make a histogram and use 30 uniformly spaced bins within $[0,1]$. We show these histograms for the validation set of RF later in Fig.~\ref{fig:RF_scores}. 
It is worth noting that although automated algorithms like \texttt{CalibratedClassifierCV} exist in the \texttt{sklearn} library, we performed our own calibration using Eq.~\eqref{eq:Probability_calibration} and show the results in the bottom panel of Fig.~\ref{fig:RF_scores}. 
%\markc{This next sentence is not apparent to me after looking at Fig. 2. Maybe I am not familiar with calibrating RF outputs, but perhaps further explanation is needed.}
We see that the RF output probabilities $p^\mathrm{ext}_\mathrm{uncalibrated}$ are sometimes overconfident: $p^\mathrm{ext}_\mathrm{uncalibrated}\sim 0.9$ should imply a factor of $\sim$9 ratio of the two histograms, but the observed ratio is smaller. In some other cases, they are underconfident (when $p^\mathrm{ext}_\mathrm{uncalibrated}\lesssim 0.4$) due to being overly conservative, potentially due to the small number of training examples.

\subsection{Deep-set neural network classifier}

Typical ML models work with a fixed-dimensional input feature vector. Deep-set is a simple generalization of DNNs which works with a flexible-dimension feature vector~\cite{Zah17_DeepSets}. Another useful property of Deep-set is that it is permutation invariant, i.e., one does not need to specify any ordering of the input feature vectors. For the RF, we had ordered the top three environmental triggers by SNR, however one could also think of doing the ordering by $\Delta t$ or template ID. For Deep-set, however, there is no assumption involved about the ordering.
%  which accounts for inputs of variable number of features. 
We estimate the probability of a particular candidate being a signal by training a Deep-set model as

\be
p'_\mathrm{NN}(\textbf{x}, \{\textbf{y}^{(0)}, \textbf{y}^{(1)}, \textbf{y}^{(2)},...\}) \equiv p_\mathrm{NN} \Big(\textbf{x}, \sum_i f_\mathrm{NN}(\textbf{y}^{(i)})\Big)\,,
\label{eq:Deepsets}
\ee
where the functions $f_\mathrm{NN}$ and $p_\mathrm{NN}$ are simple multi-linear perceptons which are trained using gradient descent by minimizing the binary cross-entropy loss given by

\be
\mathcal{L} = -\frac{1}{N}\sum_{j=1}^N p^{(j)}_\mathrm{label} \log p^{(j)}_\mathrm{NN} + (1-p^{(j)}_\mathrm{label}) \log (1-p^{(j)}_\mathrm{NN})\,,
\ee
where the label $j$ iterates over candidates. $p^{(j)}_\mathrm{label}$ is discretely either 1 or 0 for the signal and glitch candidates, respectively. We have divided the input features into two broad sets: $(i)$ $\textbf{x}$ determines the fixed-length feature set, which in our case is $\textbf{x} = \{\int\frac{|h_\mathrm{norm}(f)|^2}{\mathrm{PSD}_\mathrm{local}(f)}df, \Delta t_\mathrm{hole}, N_\mathrm{noisy\, bands}\}$. $(ii)$ $\{\textbf{y}^{(i)}\}$ is the variable-length feature set, where $i$ iterates over the set of triggers nearby to a given candidate and $\textbf{y}^{(i)} = \{\Delta t^{(i)}_\mathrm{trigger}, \mathrm{SNR}^{(i)}, c_0^{(i)}, c_1^{(i)}\}$  (see Section~\ref{sec:SummaryStats} for a definition of the input features). 
%which we also showed in the right panel of Fig.~\ref{fig:Feature_Imp}.
% \markc{perhaps it is better to explicitly mention that $\textbf{x}$ are the features on the left panel of Fig.~\ref{fig:Feature_Imp} while each $\textbf{y}_i$ is a set of the features on the right panel. Maybe even mention this earlier to motivate more explicitly why we want to use deep-set}.
Due to the summation operation in Eq.~\eqref{eq:Deepsets}, the equation is straightforward to use even if the length of the set $\{\textbf{y}^{(i)}\}$ varies for different candidates. In case there are no nearby triggers, we use a learnable tensor that serves as a placeholder. We compare the performance of the RF with the deep-set network later in Appendix~\ref{sec:Deepset_comparison} and  Fig.~\ref{fig:Deepset_comparison}. We find that the RF outperforms the deep-set classifier and therefore we use RF as our default ML model for all the results in this paper. We leave further discussion of the deep-set classifier to Appendix~\ref{sec:Deepset_comparison}.

  \begin{figure*}
    \centering
    \includegraphics[scale=0.4,keepaspectratio=true]{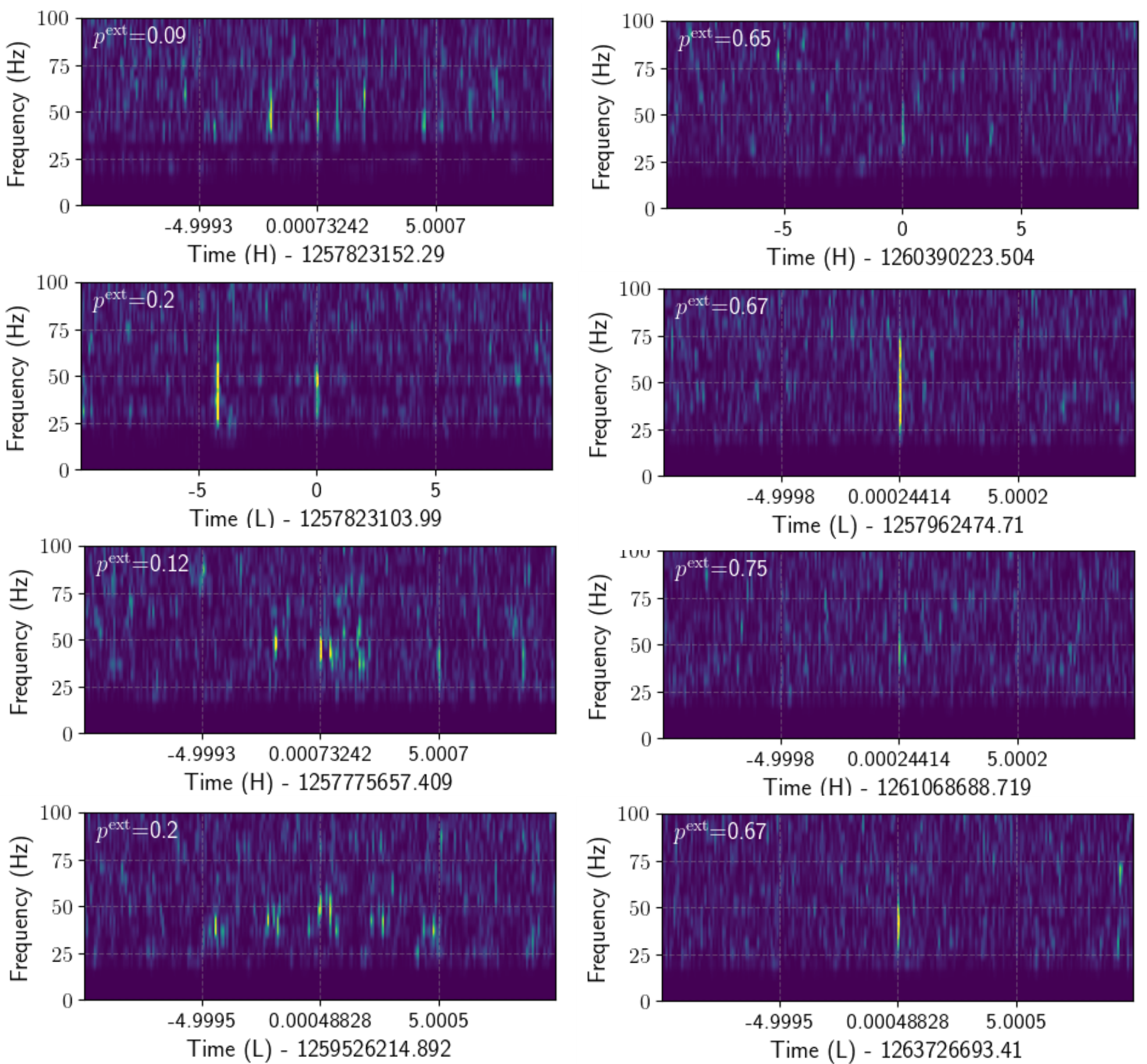}
    \caption{We show a few examples with relatively low (\emph{left}) and high (\emph{right}) values of $p^\mathrm{ext}$ (probability of the central candidate being a GW signal using only the nonlocal strain information) from the ML model.
    % Qualitatively, the $p^\mathrm{ext}$ values are consistent with a visual inspection of the data.
     The low $p^\mathrm{ext}$ value candidates are more likely to be dubious as we see that that similar triggers occur too frequently in their vicinity. Note that these are examples of the candidates which had cleared all the previous cuts/vetos in the pipeline and were only down/up-weighted by \texttt{TIER}. In some cases, \texttt{TIER} indeed provides complementary information which was missed by the traditional search pipelines.}
    \label{fig:Glitch_examples}
    \end{figure*}

\begin{figure*}
  \centering
  \includegraphics[scale=0.57,keepaspectratio=true]{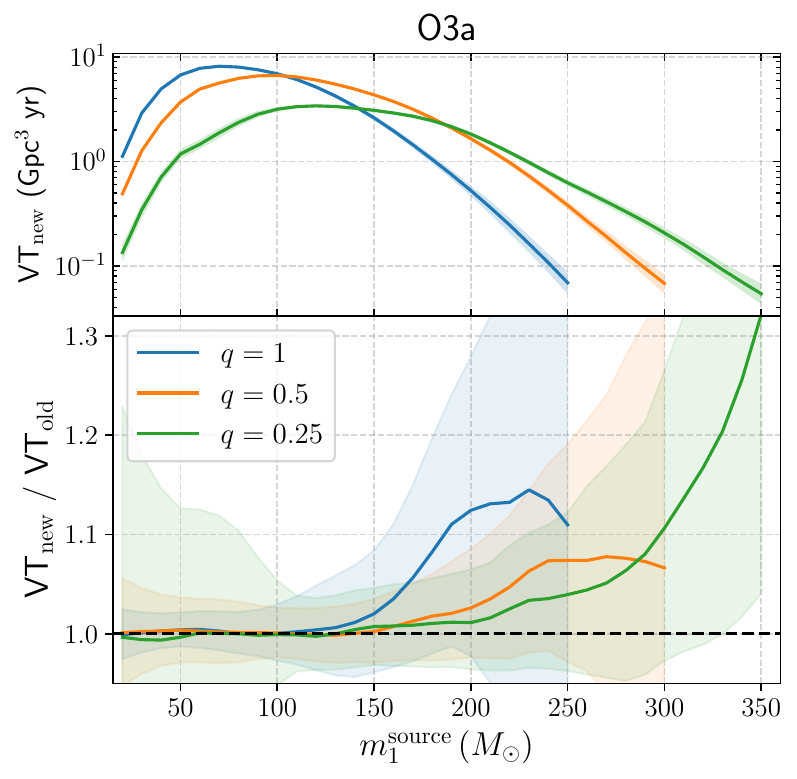}
  \includegraphics[scale=0.57,keepaspectratio=true]{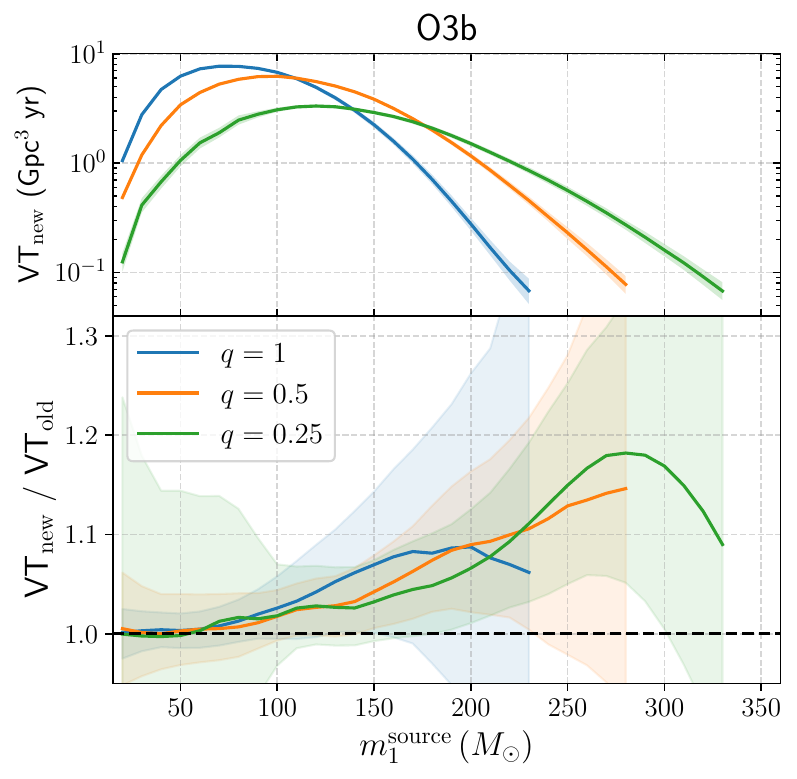}
  \caption{Increase in the sensitive volume time (VT) of the search pipeline upon using \texttt{TIER} on the data from the LIGO-Virgo-Kagra (LVK) observing runs O3a (left) and O3b (right). Note that O3a and O3b curves are different as the behavior of noise transients is different in the runs (even as the overall PSD curves for O3a and O3b are similar). The curves labelled \emph{old} are from the \texttt{IAS-HM} pipeline estimates from Ref.~\cite{Meh25_VT_HM}, while the \emph{new} curves are from the same pipeline with \texttt{TIER}. We use the injection catalog released by the LVK collaboration on Zenodo~\cite{zenodoLVK} to estimate the VT values. The shaded region is the $90\%$ confidence interval estimated by bootstrap resampling of the injection samples. The sensitivity increase is primarily concentrated in regions of high masses and asymmetric mass-ratio (these are precisely the systems with short-duration signals, and are the most plagued by transients like glitches). We also improve the significance of some of the near-threshold GW candidates detected in the \texttt{IAS-HM} search in these regions of parameter space (see Fig.~\ref{fig:m1_z}).
  %\markc{Explain the shaded region}
  }
  \label{fig:VT}
  \end{figure*}
  
\subsection{Training and test sets}
\label{sec:Training_test_sets}
As ML algorithms are susceptible to overfitting, it is important to make independent datasets for training and testing the model. In Section~\ref{sec:SummaryStats}, we discussed training our model on background triggers obtained using timeslides. We however also want to generate predictions on the same set of background triggers, as these are used in estimating the updated FARs of coincident candidates.

We show a schematic of our procedure in Fig.~\ref{fig:k-fold-training} in the Appendix. We first divide the strain data into three repeating chunks, with each chunk being 100 seconds long. 
%\MZ{What does repeating chunks mean? I don't understand the procedure.}
The chunk label for a candidate is determined by mod(int($t_\mathrm{candidate}/100$), 3), where $t_\mathrm{candidate}$ is its GPS time. To generate predictions for candidates falling within a particular chunk, we train the model on the remaining three chunks (e.g., for chunk 1 candidates, we train on chunks 0 and 2). %\MZ{Maybe be nice to those that don't know what $k$-fold cross-validation is. Can we explain it in english?} 
We repeat this process for three iterations to generate predictions for all candidates. Note that our procedure to generate predictions from the ML model is similar to the to the $k$-fold cross-validation technique used widely in ML.

Instead of splitting the data into time-chunks, an alternative procedure could be to naively split the set of background candidates randomly into training and evaluation sets (e.g. by a 70\%-30\% split) without accounting for time information. However, in the timeslides dataset, loud background triggers can occur multiple times (the same loud background trigger in Hanford can spuriously pair up with multiple time-shifted triggers in Livingston). During a random split, such a trigger can occur in both training and evaluation sets, causing contamination.

%\markc{I'm not sure if I understand the message of this section completely. So because we need to include $p^{\rm ext}$ for all the background triggers for ranking and FAR estimation, we cannot just split the dataset into training and validation set naively (e.g. by a 70-30 split)? Is that why we need to use $k$-fold cross-validation? Perhaps we should reword things so that the message is clearer.}

\section{Results}
\label{sec:results}

Once the ML model is trained, we use it to compute $p^\mathrm{ext}$ for the background and coincident candidates from the \texttt{IAS-HM} search, and also for the GW injections from the catalog described below. We then use Eq.~\eqref{eq:NeymanPearson} to calculate our new ranking statistic for the candidates by combining their $p^\mathrm{ext}$ values with the traditional ranking statistic (see~\cite{Wad23_Pipeline} for an overview of the traditional ranking statistic used in our \texttt{IAS-HM} search). This leads to a change in the the FAR and $\pastro$ (probability of being of astrophysical origin) of the candidates.

We show the performance of the ML model for one of the high-mass template banks (i.e. bank 14, which corresponds to $M_\mathrm{tot}\sim 250\, M_\odot$~\cite{Wad23_TemplateBanks}) for the Hanford O3b data as an example in Fig.~\ref{fig:RF_scores}. 
%\MZ{The way the figures appear are quite far from there they are referenced.}
In the bottom panel of Fig.~\ref{fig:RF_scores}, we show the calibration of the RF scores to obtain the Bayesian probability $p^\mathrm{ext}$ using the method described in Section~\ref{sec:Calibrating_RF_output_probability}. In the top panels, we see a relative overdensity of the GW injections (background triggers) in the region of high (low) RF scores. This shows that the ML model is able to use the extended strain to determine the nature of the candidate in particular cases. We see that the difference is larger in the low RF score region, showing that the model can reliably classify some of the candidates as glitches. However, some of the glitches also occur in clean extended regions, in which case the model cannot reliably classify them, and the model predicts $p^\mathrm{ext}$ values close to 0.5.

To qualitatively check if the $p^\mathrm{ext}$ values are consistent with a visual inspection of the data, we show a few of the high IFAR candidates with relatively large/small $p^\mathrm{ext}$ values in Fig.~\ref{fig:Glitch_examples}. 
%\markc{I think that the figures might have to be reordered} 
We indeed see that the high $p^\mathrm{ext}$ value candidates have a larger number of short-duration high SNR triggers nearby to them. Note that these candidates had cleared all the previous tests in the pipeline, and it is interesting to see that \texttt{TIER} algorithm is able to downweight some of these persistent background candidates.

To quantify the improvement in the sensitive volume-time of the search pipeline, we use the injection catalog released by the LVK collaboration on Zenodo~\cite{zenodoLVK}. Our volume-time $\overline{VT}$ computation is similar to our previous study~\cite{Meh25_VT_HM}. We use a Monte Carlo integral over the injections found in our pipeline over IFAR $>1$ year using~\cite{Tiw17_VT_estimation}:

\begin{equation}
     \overline{VT} \simeq \dfrac{T_{\mathrm{obs}}}{N_{\mathrm{draw}}} \sum_{j=1}^{N_{\mathrm{found}}} \dfrac{f(z_j)\, \dfrac{1}{1+z_j}\,\dfrac{dV_c(z_j)}{dz}\, p(\theta_j)}{{\pi_{\mathrm{draw}}(\theta_j, z_j)}}\,,
\label{eq:VT_est}
\end{equation}
where $T_{\mathrm{obs}}$ is the observation time of the dataset analyzed, $p(\theta)$ is the astrophysical probability distribution over the intrinsic source parameters $\theta = (m_1^{\mathrm{s}}, m_2^{\mathrm{s}}, \Vec{s}_1, \Vec{s}_2)$, $V_c$ is the comoving volume and $z$ is the redshift, $N_{\mathrm{draw}}$ is the number of injections generated from the sampling distribution, and
$\pi_{\mathrm{draw}}$ is the sampling probability of the injections.
We show the $\overline{VT}$ results in Fig.~\ref{fig:VT}. We see an improvement because there are more injections detected with IFAR $\geq 1$ year after application of \texttt{TIER}. The improvements are concentrated in regions of high-masses and asymmetric mass-ratio. These are precisely the systems with short-duration signals, and are the most plagued by transients like glitches. Note that to reduce the discreteness effects due to a finite number of injections, we make bins in $m_1$ and $q$ and average over the injections within each bin. In particular, each point on the VT curves in Fig.~\ref{fig:VT} corresponds to using an astrophysical prior with the mean at the center of the bin and standard deviation of $\sigma^2=0.1$:
\begin{align}
  P_{m_{1s},q}(m_{1s}^\prime, q') \sim\,  &{\rm Lognormal}(\mu=m_{1s}, \sigma^2)\\ &\times  {\rm Lognormal}(\mu=m_{1s}q, \sigma^2)\,.
\end{align}
We also show the uncertainty in $\overline{VT}$ evaluated using the bootstrap method, i.e., resampling the injections with replacement~\cite{Meh25_VT_HM}. The uncertainty is higher at higher masses and lower mass ratios, as the number of effective injection samples is lower. We also truncate the curves for different $q$ at a lower limit of $\overline{VT}=0.05$ Gpc$^3$ yr to avoid points with huge error bars.

%%%%%%%%%%%%%%%%%%%%%%%%%%%%%%%%%%%%%%%%%%%%%%%%%%%%
\section{Discussion}
\label{sec:discussion}

Once the RF model is trained, one can determine which of the input summary statistics of the extended data are most useful for its prediction. We show this feature importance plot in Fig.~\ref{fig:Feature_Imp}. This could also help in designing a theoretical model using the most informative features. We see that the number of triggers is not a good predictor of the significance of a candidate. Instead, a few of the loudest triggers nearby to the candidate provide useful information about the candidate's significance. The next most important parameters are related to the local PSD, and the time difference of the candidate from very loud disturbances in the data. 

\begin{figure}[!h]
  \centering
  \includegraphics[width=0.47\textwidth]{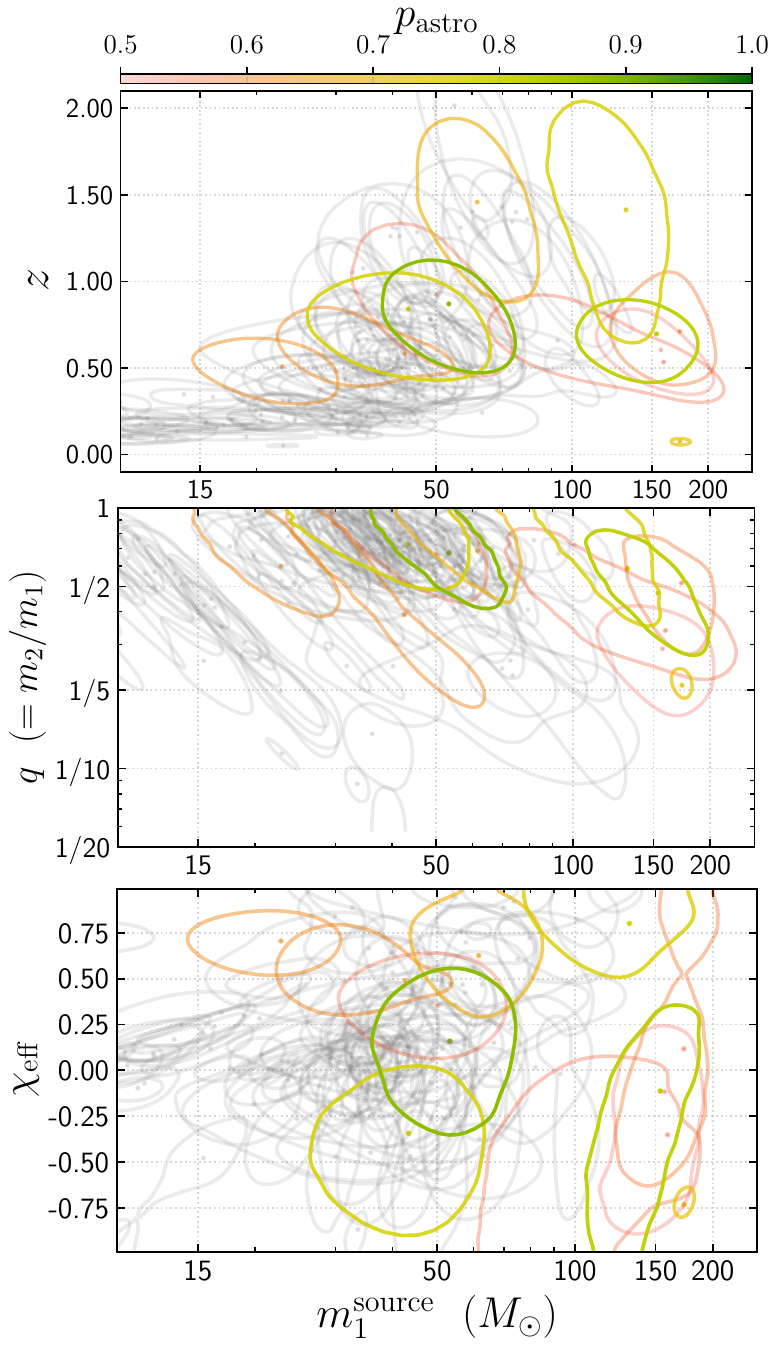}
  \caption{Updated version of the new candidate events from the \texttt{IAS-HM} search in \citet{Wad23_HM_Events} after the application of the \texttt{TIER} algorithm (these plots are also reported in the updated version of Ref.~\cite{Wad23_HM_Events}). The relative change in $\pastro$ and IFAR values of the new candidates upon application of \texttt{TIER} is quantified in Fig.~\ref{fig:pastro_comparison}. 
  We show the source-frame primary mass ($m_1$), redshift ($z$), mass ratio ($q$) and effective spin ($\chieff$). In colored contours, we show properties of the new candidate events found in our O3 search with $\pastro >0.5$ ($\pastro$ is the probability of events being of astrophysical origin, as opposed to being a noise transient). The color of the contours corresponds to the $\pastro$ values of the events. All the previously reported events from O1-O3 runs are shown in transparent gray contours~\cite{O1catalog_LVC2016, ias_pipeline_o1_catalog_new_search_prd2019, gwtc1_o2catalog_LVC2018, ias_o2_pipeline_new_events_prd2020, lvc_gwtc3_o3_ab_catalog_2021, Kol24_Ares_ML_Search,lvc_o3a_deep_gwtc2_1_update_2021, Ols22_ias_o3a}.
  %detected so far from all the search pipelines across the first three observing runs of LVK.
The posterior contours are obtained from parameter estimation runs and enclose $50\%$ of the probability, and median values are represented by dots. Interestingly, some of the new candidate events have support in the ranges corresponding to IMBH masses, pair-instability mass gap, high redshift, low mass ratio, and negative effective spin.
%\markc{The colors of the contour lines seem not to match the color bar, perhaps because of the alpha you are using. Maybe you can use alpha closer to 1 for the colored contours, so they look closer to the colors in the colorbar, and they can also standout more from the gray contours.}
}
    \label{fig:m1_z}
\end{figure}

The results for coincident candidates in the LVK O3 data upon the application of \texttt{TIER} are given in an updated version of the \texttt{IAS-HM} detection catalog in Ref.~\cite{Wad23_HM_Events}. In this paper, we replicate an updated version of their plot in Fig.~\ref{fig:m1_z}. We show properties of the new candidate events found by the \texttt{IAS-HM} pipeline in colored contours.
The contours correspond to the posterior of properties of the signals and are obtained using parameter estimation runs as described in~\cite{Wad23_HM_Events}. For the new candidate events, we quantify the relative change in $\pastro$ and IFAR values of the new candidates upon application of \texttt{TIER} in Fig.~\ref{fig:pastro_comparison} in the Appendix. %\markc{I think we can think of showing the improvements in the plot, or in a separate plot / table, especially when we mentioned these improvements in the abstract. Now the plot only shows the updated $p_{\rm astro}$ but not its improvements.} 
For comparison, we show in gray all the previously reported events: the GWTC-3 LVK catalog, the OGC-4 catalog, the ARES catalog, and the previous IAS (2,2)-only catalogs.

% Apart from an improvement in the sensitive volume-time, \texttt{TIER} also leads to a change in the the IFAR and $\pastro$ of previously reported candidate events. We provide a new catalog in the updated arXiv version of the paper~\cite{Wad23_HM_Events}

% Our method does not involve any knowledge of the astrophysical model other than the fact that physical events occur randomly in time. Furthermore, we do not need expensive injection campaigns to generate the training data. Our lean and modular framework is easy to train and can be used with triggers from any search pipeline.

% Tests for signal consistency are done by dividing the signal into multiple parts and then checking the consistency of each part separately.
% \be
% \tilde{\rho}_\mathrm{rescaled} = \rho \left[\frac{1}{2} \left(1+\Big(\frac{\chi^2}{n_\mathrm{chunks}}\Big)^3\right)\right]^{-1/6}
% \ee
% However, the efficacy of this test starts to decrease for short-duration waveforms as making independent chunks becomes tougher. 

\subsection{Future work}
\label{sec:Future_work}

By using \texttt{TIER}, we have a more accurate significance of candidates in the high-mass and asymmetric mass-ratio parameter regimes. Finding new high-mass candidate events is interesting as they can help us answer open questions about pair-instability supernova physics and for studying formation channels of IMBHs. Having more sensitivity in the asymmetric mass-ratio regime is useful as it will help us disentangle the formation channels of BHs. We plan to perform population studies using our new catalog and the new search selection function in an upcoming work~\cite{Meh25_Population_HM}.

% \subsubsection{Including environmental channels}
In the \texttt{TIER} framework, we included various features of the environmental strain in the search ranking statistic.
One could also include features from auxiliary channels as an input to the ML model~\cite{God20_iDQ_GSTLAL, Dav22_iDQ_PyCBC, God20_iDQ_GSTLAL}. We leave this point to an upcoming study.

In Eq.~\eqref{eq:NeymanPearson}, we broke down the ranking statistic into two independent parts, assuming that the information from the extended strain data (encoded in $p^\mathrm{ext}$) is independent of the information from the local data. Another way of incorporating $p^\mathrm{ext}$ inside a search pipeline could be in background-based reweighting of triggers, wherein we penalize candidates differently based on the their probability of occuring due to noise (see Section~III B of~\cite{Wad23_Pipeline}). For example, we penalize triggers coming from shorter templates (e.g., those corresponding to negative $\chieff$ or asymmetric mass-ratios) more strongly compared to the triggers coming from longer templates (as the shorter templates are more prone to glitches~\cite{Wad23_Pipeline,LIGO_O1, CoherentScore, psd_drift, BlipGlitches}).

One could further extend this methodology to do background reweighting differently for triggers with different $p^\mathrm{ext}$ values. For example, we can empirically determine if there is a possible correlation between the SNR of a trigger being high and its $p^\mathrm{ext}$ value. This information can then be used to increase the sensitivity of the search pipeline. 
%The downside of this approach compared to using Eq.~\eqref{eq:NeymanPearson} is that it couples the information from the environmental features with the information coming from the search pipeline (thus one cannot simply multiply the two terms as in Eq.~\eqref{eq:NeymanPearson}). 
We leave exploring this direction to an upcoming study. Futhermore, \texttt{TIER} could also be used to aid gating or inpainting of loud noise transients or flagging poor data quality regions~\cite{Ess20_iDQ,Dav22_iDQ_PyCBC}. 

\subsection{Comparison to previous literature}

A few other search pipelines have included some features of the extended strain data to determine the significance of candidates.
The local rate of noisy triggers is one of the parameters included in the ranking statistic of the GstLAL pipeline~\cite{Can15_RankingStat_GSTLAL, Tsu23_GSTLAL}. The ML-based search method ARES~\cite{Kol24_Ares_ML_Search} also uses a local rate of noisy triggers to divide triggers into sub-categories and rank them separately. We however see in our feature importance plot in Fig.~\ref{fig:Feature_Imp} that the total number of triggers is not a good predictor. Instead, properties of the triggers (e.g., their SNR) nearby the candidate provide more useful information about the candidate's significance.

\section{Conclusions and Summary}
\label{sec:conclusion}

To determine the significance of a GW candidate, search pipelines typically only use strain data information in the vicinity of the candidate signal. However, extended strain data also carries valuable information which can be used to more accurately determine the significance of a candidate. We develop a machine learning (ML) classifier based framework called \texttt{TIER} (Trigger inference using Extended strain Representation) to extract this information (Fig.~\ref{fig:TIER}). To make our ML classifier easy to train, we use sparse summary statistics of the extended strain data (Fig.~\ref{fig:Feature_Imp}) to train the model. We see that our ML model can distinguish between background triggers and GW signals using information from the extended strain data (Figs.~\ref{fig:RF_scores}, ~\ref{fig:Glitch_examples}). The output of the ML models can be easily integrated into ranking statistics of search pipelines.  We also show that using \texttt{TIER} can improve the sensitivity of the \texttt{IAS-HM} search pipeline by up to $\sim$20\% in the sensitive volume-time (Fig.~\ref{fig:VT}), with improvements being largely concentrated in regions of high-mass and asymmetric mass-ratio systems. This leads to an increase in significance of a number of near-threshold candidates previously detected in these regions of parameter space (Fig.~\ref{fig:m1_z}).

We provide code modules associated with the \texttt{TIER} framework in a jupyter notebook at \url{https://github.com/JayWadekar/TIER_GW}. We also provide data products for a particular bank (bank 14) which can be used to reproduce some of the plots in this paper. We plan to release the full catalog of background and foreground triggers and the extended-strain dataset for the entire region of the search parameter space along with our companion paper (Ref.~\cite{Pim25_Env_ML}).

\acknowledgments

We thank Shivam Pandey, Aaron Zimmerman, Derek Davis, Muhammed Saleem, and Archana Pai for helpful discussions. We also thank ICERM for their hospitality during
the completion of a part of this work. ICERM is supported by the National Science Foundation under Grant No. DMS-1929284. MZ acknowledges support from the National Science Foundation NSF-BSF 2207583 and NSF 2209991 and the Nelson Center for Collaborative Research.
T.V acknowledges support from NSF grants 2012086 and 2309360, the Alfred P. Sloan Foundation through grant number FG-2023-20470, and the BSF through award number 2022136.
%We thank the anonymous referee for their critical comments and various useful suggestions which improved the manuscript.

This research has made use of data, software and/or web tools obtained from the Gravitational Wave Open Science Center (\url{https://www.gw-openscience.org/}), a service of LIGO Laboratory, the LIGO Scientific Collaboration and the Virgo Collaboration. LIGO Laboratory and Advanced LIGO are funded by the United States National Science Foundation (NSF) as well as the Science and Technology Facilities Council (STFC) of the United Kingdom, the Max-Planck-Society (MPS), and the State of Niedersachsen/Germany for support of the construction of Advanced LIGO and construction and operation of the GEO600 detector. Additional support for Advanced LIGO was provided by the Australian Research Council. Virgo is funded, through the European Gravitational Observatory (EGO), by the French Centre National de Recherche Scientifique (CNRS), the Italian Istituto Nazionale di Fisica Nucleare (INFN) and the Dutch Nikhef, with contributions by institutions from Belgium, Germany, Greece, Hungary, Ireland, Japan, Monaco, Poland, Portugal, Spain.

\appendix

\section{Comparison of the results from RF and deepset network}
\label{sec:Deepset_comparison}

\begin{figure}
  \centering
  \includegraphics[scale=0.7,keepaspectratio=true]{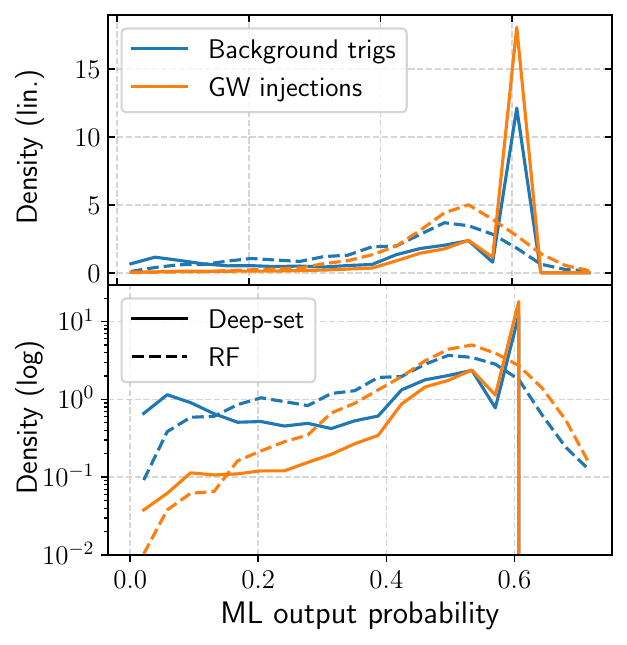}
  \caption{Similar to Fig.~\ref{fig:RF_scores} in the main text, but comparing the performance of RF (dashed) with deep-set classifier (solid). We show the histograms of the rescaled probabilities from the two ML models. We see that there is a relative overdensity of the GW injections in the high score region for both models, and vice versa for the background triggers. However, the performance difference between the two models is not substantial (see also Appendix~\ref{sec:Deepset_comparison}). We prefer to use RF for the main results as it is much easier to train due to substantially lower computational cost.}
  \label{fig:Deepset_comparison}
  \end{figure}

As described earlier in Section~\ref{sec:ML_models} in the main text, we trained a RF model and a deep-set network to classify the candidates as signals or noise transients. We show a comparison of the performance of the two models in Fig.~\ref{fig:Deepset_comparison}. 
The deep-set classifier does not show a substantial improvement over the RF model at distinguishing background triggers from GW signals.
The mean binary cross-entropy loss is 0.6364 for the deepset model, in comparison to 0.6367 for RF.
Although performances of the two methods are similar, training the deep-set network is comparatively much more expensive compared to the RF model. 
%Furthermore, finding the optimal hyperparameters for the deep-set for different template banks is also challenging. 
Hence, we use RF as our default ML model for all the results in the main text.

\section{Extra figures}

\begin{figure}
  \centering
  \includegraphics[scale=0.7,keepaspectratio=true]{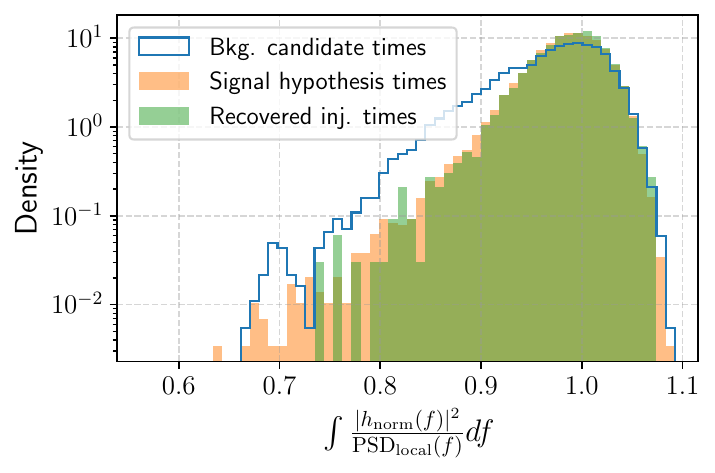}
  \includegraphics[scale=0.7,keepaspectratio=true]{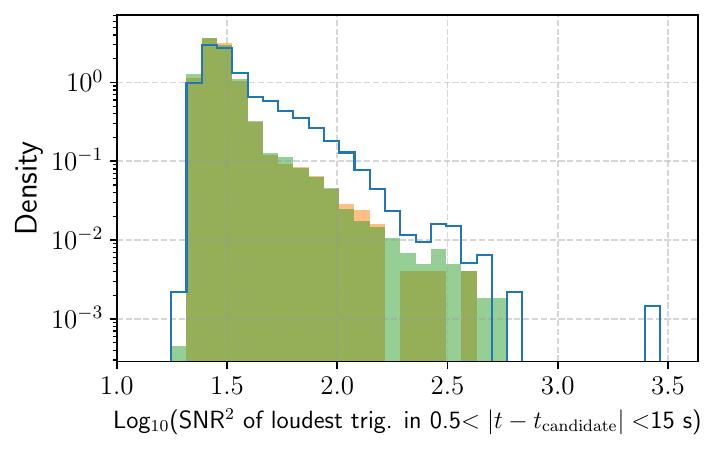}
  \caption{We show the distribution of two most important summary statistics of the extended strain data (see section~\ref{sec:data} and Fig.~\ref{fig:Feature_Imp}). The top panel corresponds to the local sensitivity of the data (as determined by the local PSD) at the location of the trigger times. The bottom panels shows the log$_{10}$(SNR$^2$) distribution of the loudest trigger in the vicinity of the candidates. We show the times of the background candidates collected under the noise hypothesis, and the set of simulated times collected under the signal hypothesis. These two sets were used for training the ML model as discussed in section~\ref{sec:Bkg_signal_candidates}. We also show for comparison the distribution of times of the recovered injections from section~\ref{sec:results} which were used for testing the model. The properties of the extended strain data at the times of background candidates is significantly different from the other two cases.}
  \label{fig:PSD_local_distribution}
  \end{figure}

\begin{figure}
  \centering
  \includegraphics[width=0.45\textwidth]{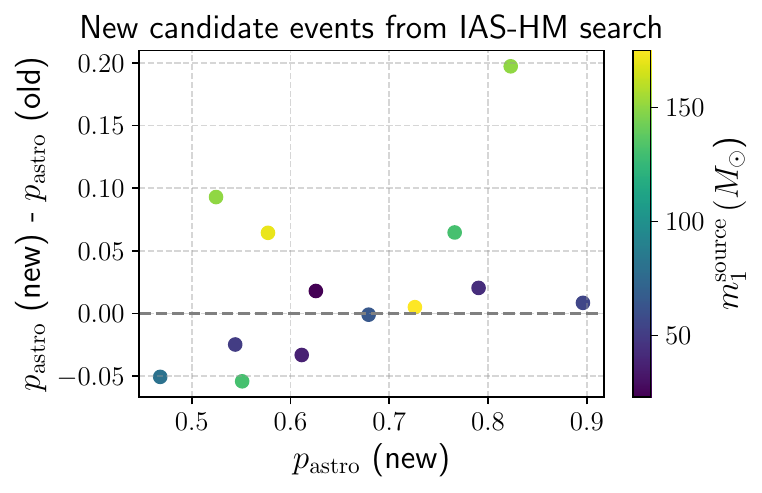}
  \includegraphics[width=0.45\textwidth]{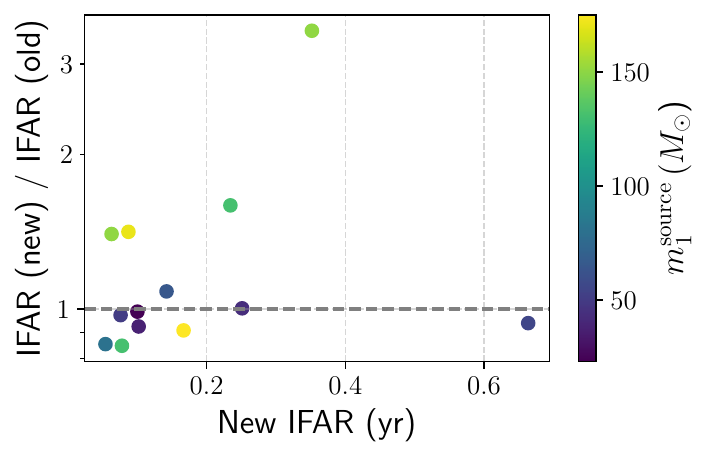}
  \caption{We compare the change in $\pastro$ and IFAR of the new candidate events found in the \texttt{IAS-HM} search~\cite{Wad23_HM_Events} upon the application of the \texttt{TIER} algorithm.
  We see both positive and negative fluctuations in IFAR and $\pastro$ values. This is expected, as the new candidate events are at the edge of the detection threshold. Some of the new candidate events were probabilistically likely to be glitches and the \texttt{TIER} algorithm is able to downweight those.}
  \label{fig:pastro_comparison}
  \end{figure}

\begin{figure}
  \centering
  \includegraphics[scale=0.25,keepaspectratio=true]{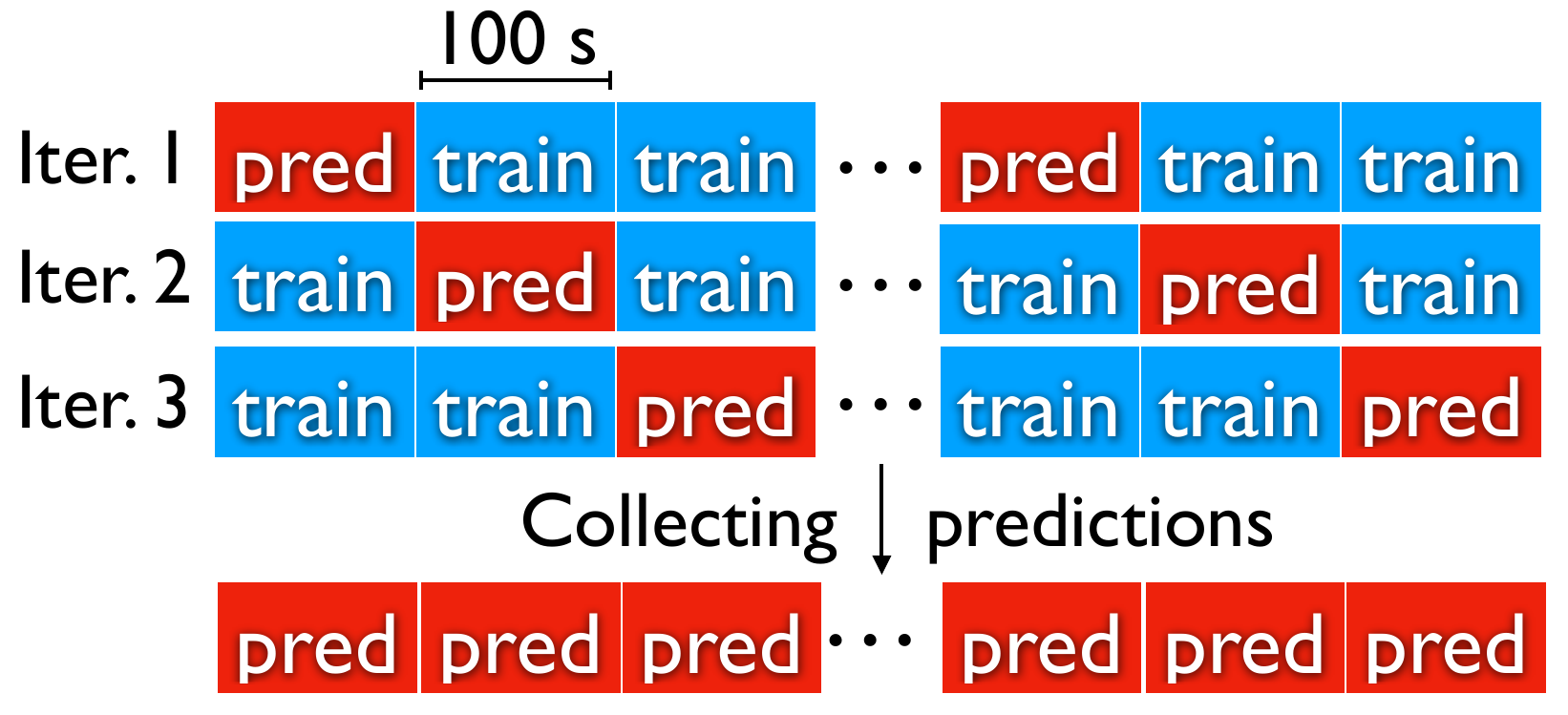}
  \caption{We described in section~\ref{sec:Training_test_sets} how to make training and testing datasets for ML models. It is important to make independent datasets for training and testing ML models, as they are susceptible to overfitting. In our case, the same background triggers are used for both training and testing the model. Hence we divide the dataset into three repeating chunks and obtain predictions from the model in three iterations. Our method ensures that ML predictions for each candidate come from a model that was not trained on the time-chunk that contains the candidate.}
  \label{fig:k-fold-training}
  \end{figure}

% \begin{figure}
% \centering
% \includegraphics[scale=0.6,keepaspectratio=true]{Probability_calibration_2.pdf}
% \caption{Calibrated output probabilities from the random forest. The raw curve is obtained from Eq.~\eqref{eq:Probability_calibration} and we also show the monotonic fit which is then used in our ranking statistic.}
% \label{fig:Probability_calibration}
% \end{figure}

%%%%%%%%%%%%%%%%%%%%%%%%%%%%%%%%%%%%%%
% \bibliographystyle{apsrev4-1}

\clearpage
\bibliographystyle{apsrev4-1-etal}
\bibliography{GW, GW_ML}
\end{document}